# POST-FUSION MONOLITHIC HYBRID RECOMMENDER SYSTEM FOR SUGGESTING RELEVANT MOVIES TO USERS

**Mahdi Rezapour**

## ABSTRACT

Background Recommendation systems are essential tools for facilitating user access to information and addressing the challenge of information overload. They operate by analyzing historical behaviors to learn user preferences. This study explores a hybrid recommendation approach that combines collaborative filtering methods: analyzing sequences of watched movies and leveraging movie ratings. The goal is to enhance recommendation accuracy by dynamically adjusting these methods based on specific use cases. The proposed method uses a weighted post-fusion technique to combine the two collaborative filtering approaches. For instance, potentially higher weights are assigned to the rating matrix when the majority of data includes ratings, while sequential data receives more emphasis when ratings are scarce. Technique could integrate sequential data with ratings effectively, capturing both short-term and long-term user preferences. The approach addresses limitations of standalone models, such as insufficient long-term preferences or lack of rating considerations. Integrating heterogeneous data sources within a unified recommendation framework leads to richer, more adaptive user modeling. The dynamic fusion approach not only personalizes results but also aligns recommendations with evolving user patterns. Future work may include learning adaptive weights automatically through meta-learning, incorporating metadata like genres or timestamps, and scaling to real-world datasets with millions of users and items. This hybrid methodology highlights the importance of combining multiple data sources to create more personalized and effective recommendations. The study provides a framework for adapting weights to align with varied use cases, paving the way for advancements in hybrid recommendation systems.

### 0.1 Introduction

The internet contains a large amount of information which needs to be filtered to determine the suitability of the contents for certain users. Given the vast amount of information, users are often faced with an enormous number of products and recommended items. Recommender system (RS) has become crucial in information filtering for providing most reliable recommendations. As part of the information filtering system, they help users find relevant information from a large amount of information. Recent developments in information retrieval and recommendation systems provide people with access to a large amount of information quickly and efficiently.

Recommendation system has been employed in many domains from tourism [23] [17], e-commerce [22] to music [**?** ]. Amazon, for instance, was one of the initial companies, patented the first version of the recommender system as early as 2004, which increased the profit by 29% [11]. Netflix, on the other hand, is another example which has been using recommenders' systems to decrease the number of canceled subscriptions and to increase the average time of users interacting with the application (i.e., number of streaming hours). It was anticipated that a combination of various recommendation systems such as "trending now", "because you watched", and "continue watching" have saved up to one billion dollars a year for that company [3]. So, it is evident that recommendation systems play a pivotal role in the world of online and advertisement.

Recommender systems could be used as tools for information filtering to deal with such problem, while providing users with more personalized and relevant information [13] [12]. To give recommendations, systems require to analyze the past preference of the users with similar interests.

Deep learning (DL) techniques have been used in various domains such as speech recognition, computer vision and natural language processing. While DL approaches have been used in preference prediction tasks [26],those techniques could be used for capturing non-linear user-item interaction and latent data in the higher dimensions. Various techniques could be used for a means of recommendation systems from neural-based network [29] [19] to advanced graph neural network [24] [28].

There are mainly three types of recommendation systems including collaborative filtering (CF), content-based (CB) filtering and hybrid recommendation system [14]. CB is a method for recommending items with attributes, similar to those that users liked in the past [15]. CF is one of the popular recommendation techniques and a significant number of recommender's systems have been developed based on this technique due to its power in overcoming the limitations of the CB technique [16-18].

A hybrid recommendation system, on the other hand, has been proposed to solve the limitation of content-based filtering, which is relied on metadata of the users' items and collaborative filtering, being relies on users item rating data [10]. Hybrid system, which mainly combine CF and CB, could be divided into weighted, where the weights are gradually adjusted based on to the degree to which the user's evaluation of an item matches the evaluation predicted by the recommendation system, or switching, where the recommendations would be proposed based on situation, or cascaded hybridization, where initially one of the recommendation model will be used for creating a candidate set with a similar taste to the users, then, another model would be used to sort the candidate in the order of items most suited for users taste [20].

On the other hand, there are three main strategies for a hybrid recommendation system [8],including pre-fusion, middle fusion and post-fusion. Where prefusion refers to the fusion of multiple recommendations for construction of a recommendation model by combining them into a unified model and performing a feature extraction training model, and then generating recommendation results based on the fusion model. Middle fusion, on the other hand, is based on use of the recommendation algorithm as the framework, while fusing another recommender system. Lastly, post-fusion is involved with training recommendation system separately to generate recommendation results and then combining the recommendation results of each recommendation model. In this study explicit feedback of the users' data was used. As in this study, we considered two techniques of rating and sequential nature of the datasets, the next few paragraphs elaborate on those techniques.

In recommendation system like movies system where the rating dataset is used, first users rate items and then the system makes prediction about the users' ratings for items that the users has not been rated, for instance look at [30]. [29], which conducted rating prediction and achieved a promising performance by generalization of matrix factorization to neural collaborative filtering. [25], and [30], on the other hand, adopted attention network in deep neural network (NN) to find more representative features. In that study, DNN learns the latent features of users and movies as input and used that info to predict the rating score through forward propagation.

[24] used model-based collaborative filtering by utilizing the ratings of the user-item matrix dataset to generate predictions. In that study, the two widely used techniques including biased matrix factorization and a regular matrix factorization with stochastic gradient descent (SGD) were used.

Twofold objectives were discussed in the past study, where it first derive implicit ratings by employment of the CF for online transaction, even when no explicit rating info was available, and secondly to integrate the CF with sequential pattern analysis (SPA) for enhancement of the recommendation quality [15]. In that study, it was discussed that the hybrid approach of CF and SPA is better than the individual ones.

Although rating-based methods in CF could explicitly model users and items based on the rating patterns, they could suffer from sparsity issues, being related to the user-generated review, which provides rich semantic information, where the sparsity problem of rating data could be alleviated.

The sequential nature of the items that users interacted with are a key feature of many recommenders' systems as they take into consideration the sequential nature of the users' activity and are based on their recent activities. The goal of sequential recommender system is to model users' behaviors, while considering the users' experience. However, capturing that pattern is challenging as the sequence of the input is expected to grow exponentially.

### 0.2 Material and method

In this work, we employed 'self-attention' concept elaborated by [2]. Similarly, studies conducted considering that feature of the dataset. For instance, the previous study used a hybrid technique, taking into consideration both sequential patterns, and also used a unified similarity models of collaborative and content-based to find accurate relationships between items [31].The technique was found to be highly efficient in uncovering semantic and syntactic patterns across words in a sentence. Application of self-attention will help to draw context from all actions taken by users in the past. In the sequential recommendation system, a user's action sequence $S^u = (S^u_1, S^u_2, \ldots, S^u_{|S^u|-1})$ was considered into a fixed-length sequence, $s=(s_1, s_2, \ldots, s_n)$, where *n* is the length that we set our model to process [27]. Similar to that work [27], in this study, if the sequence length is less than *n*, we will add a padding until the length becomes *n*.

The author, in the past study, proposed a positional embedding for the sentence order information [2]. Similarly, as the order also exists in users' behaviors' sequences, the 'position' needs to be added as an input feature to account for that. As self-attention does not account for position of previous items, as discussed in the previous work [27], a learnable position embedding, $P \in \mathbb{R}^{n \times d}$, is injected into the input embedding as follows:

$$\widehat{M} = \begin{bmatrix} M_{s1} + P_1 \\ M_{s2} + P_2 \\ \cdots \\ M_{sn} + P_n \end{bmatrix} \quad (1)$$

$M \in \mathbb{R}^{|I| \times d}$ is the embedding matrix, and where *d* is the latent dimensionality. To account for non-linearity in the model and to consider interactions across latent dimensions, we used rectified linear unit (ReLU) activation function. On the other hand, to alleviate the overfitting issues, the dropout regularization technique was employed [18]. As the core of embedding matrix is self-attention, it is worth briefly discussing its characteristics. Self-attention, or transformers, is based on the scaled dot-product attention, defined as follows [2]:

$$Attention\ (Q, K, V) = softmax\left(\frac{QK^{\top}}{\sqrt{d}}\right)V \quad (2)$$

Where *Q* represents a query, *K* is the matrix of keys and *V is* the matrix of all values. Here the self-attention operation takes the embedding of items as an input and converts them into three matrices through linear operation.

It should be noted that in the self-attention application, all representations for heads are created from the same inputs. The initial embedding would be decomposed into $h \times d_{head}$, and the computation would be done independently.

$$MultiHead\ (Q, K, V) = Concat(head_1, \ldots, head_h)W^o \quad 3$$

$$\text{Where } head_i = Attention(QW^Q_i, KW^K_i, VW^V_i) \quad 4$$

So, each attention head would be implemented independently and, for instance, worker 1 uses $Q_1, K_1, V_1$, while worker 2 uses $Q_2, K_2, V_2$.

Here embedding is a low dimensional hidden factors accounting for users and movies vectors. The work took advantage of the matrix factorization (MF) concept, stating that the behaviors of users could be based on intrinsic hidden factors or embeddings. As sequential recommendation techniques are mainly based on self-attention, it is worth discussing some literature discussing its implications.

As this study employed post fusion, it is worth discussing its mathematical formulation of the process. There are generally 3 ways of integrating the recommendations into the two independent techniques.

# 1 Weighted Sum Fusion (Simple Approach)

Let:

- $y_{seq}$ is the output (prediction) from the NN that processes the sequential watched movies data.
- $y_{rating}$ the output from the NN that processes the rating data.

We combine these predictions linearly using weights that reflect how much confidence or influence you want to give to each source. In its simplest form, the final prediction $y_{rating}$ can be expressed as:

$$y_{Final} = W_{Seq} \cdot y_{Seq} + W_{Rating} \cdot y_{Rating} \quad 5$$

subject to:

$$W_{Seq} + W_{Rating} = 1 \quad 6$$

For instance, if the rating data is deemed more indicative of the target outcome, you might choose:

$W_{Rating}$ =0.7 and $W_{Seq}$ =0.3.w.

This way, the rating network's output has a proportionally higher influence on the final prediction.

## 2 Extended Fusion with Bias (Linear Transformation)

If we need to apply a bit more flexibility than a strict weighted average, we can include an additional bias term. This approach can capture any shift between the combined features and the desired output. The equation then becomes:

$$y_{Final} = W_{Seq} \cdot y_{Seq} + W_{Rating} \cdot y_{Rating} + b \quad 7$$

where b is a bias term. This model can sometimes be preferable if your NN outputs are calibrated differently or if you want to allow the fusion layer to offset systematic prediction errors.

## 3 Learned Fusion via a Fusion Layer

A more flexible strategy is to set up a small post-fusion network that learns how to combine the two predictions. In this case, you first concatenate the two outputs and then pass them through a dense layer:

$$y_{final} = \varphi\left(W \begin{bmatrix} y_{rating} \\ y_{seq} \end{bmatrix} + b\right) \quad (7)$$

Here,

is a weight matrix of size 1×2 (or larger to have more capacity),

is a bias term, and () is a chosen activation function (which could be linear if you prefer a weighted sum, or non-linear for more complex interactions). In this learned fusion approach, the network can automatically adjust the effective weight it gives to each prediction during training. For example, if the ratings are consistently more informative, the training process may learn a higher weight for $y_{rating}$.

In this study we employed the first approach.

The past study used self-attention mechanism for making prediction based on relatively few actions [27]. That technique seeks to find which items are relevant based on user's action history. It was discussed that as self-attention network (SAN) is inadequate in distinguishing users' long-term preference and short-term demands, a multi-layer long- and short-term self-attention network (LSSA) could be used to model sequential recommendation [4]. The technique works by first splitting the entire sequence of a user into multiple sub-sequence based on their timespan. Then, the first self-attention layer learns the users' short-term dynamics and the second one captures the user's long-term preferences through the previous subsequences and the last one.

In another study, a variational autoencoder (VAE) was employed in a self-attention network for capturing latent user preferences. In that study, self-attention vector was represented as density via variational inference. In addition, self-attention networks for learning the inference process and generative process of VAE, for capturing long-range and local dependencies were employed [14].

Considering weights, recommendation systems could be divided into weighted hybridization, where the weights are gradually adjusted based on the degree to which the user's evaluation of an item coincide with the evaluation predicted by the recommendation system, or switching hybridization, where the method changes the recommendation system based on the situation [21]. Cascaded hybridization is another weighted recommendation system, where initially one of the recommendation model for creating a candidate set with a similar taste would be set for the users and then that will be used to sort the candidate in the order of items most suited for users taste [20]. However, it should be noted that it is challenging to find out about demographic information of the users. After considering both technique and applying the weights matrix, the recommendations would be modified to correspond to the users' preference as much as possible [16]. Our approach could be implemented by considering weights across the implemented techniques, so it is worth discussing some of the past studies that employed that technique in the context of recommendation system.

[35] proposed a hybrid approach where a weighted combination of user-based, and item-based filtering were used for finding the unknown rating of an item. Utility matrix was created which maps the users rating [1]. For that matrix, the rows represent the users and columns represent as the items. There are two ways of populating that utility matrix, where the users rate a movie, or if a user watches a movie, where it is assumed, they like the movies. That matrix is usually having Boolean values. The purpose of a recommendation system is to predict missing elements. Although it is not very practical, in this study the dataset incorporates the explicit rating.

In this study, two measures of accuracy and the relevance of the recommendation system were considered. The root means square error (RMSE) was used for measuring the accuracy and the precision@K and recall@k were used for the purpose of relevance. recall@k measures the proportion of relevant items found in the top-k recommendations while precision@k measures the proportion of recommended items found in the top-k set.

Finally, a few paragraphs in appendix discuss and elaborate on the methodological approach used for implementation of the two techniques. It should be noted even though extensive efforts have been given to various hybrid recommendation techniques, especially combining CB and CF, not many studies conducted considering the two types of CF, based on sequential nature of the data, and based on its rating. For instance, combination of hybrid technique like considering demographic filtering with CF and CBF [9] or weighting method for combining user-tag, user-user and user-item CF relation in social media were used where it uses the final rating score of an item for users as a linear combination of the three CF relation [5].The appendix will discuss the methodological approaches of the two techniques of CF based on sequential nature of the dataset and also based on rating. Figure 1 is also included to provide the process of using various weights to come up with a final recommendation.

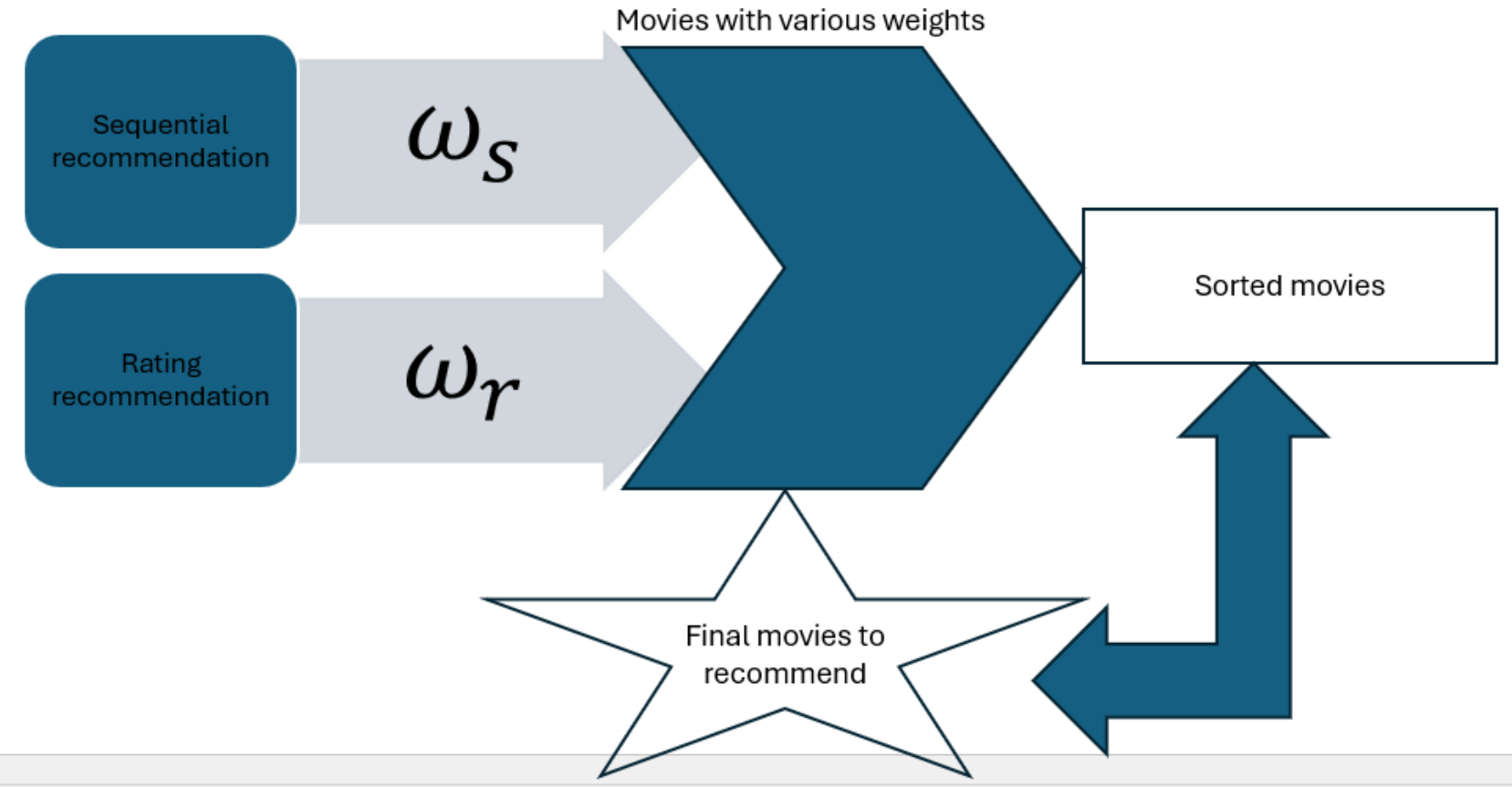

Figure 1: A sample methodological movie recommendation approach

In this study, two publicly available files of rating, which include users' ratings for various movies, and MovieLens were used. The rating file includes features like userID, movieID, and ratings. It should be noted, here users' preferences for items are called rating.

MovieLens dataset, on the other hand, is a dataset of personalize ratings of a various movie from a large numbers of users being used [6]. The dataset contains 20,000,263 ratings made by 138,493 unique users on 26,744 movies. The rating data includes columns of user_id, movie_id, rating and timestamp. More information about the used dataset are available in online resources.

#### 3.0.1 Result

In this study two recommendation techniques were used, where based on expertise, various techniques of post-fusion, for instance, could be used to consider a combination of both approaches. The sequential technique has the advantage of considering a series and sequence of watched movies, while it has the disadvantage of considering movies with potential low rating by disregarding ratings column.

Here in Table 1 and 2, we included examples of users with ID of 110007 and 65543. Again, we did not do any post processing due to the lack of knowledge in the field and the fact that there were no overlapped observations across the two techniques across the considered examples, and without overlapped most of the selected movies might end up with a single technique. In Table 1 and 2, the top row of those tables highlights the recommended movies based on ratings while the second row is based on sequential nature of the movies.

In the current study, for sequential analysis for 5 epochs, we found a loss of 6.56 and perplexity (PPL) of 705.60. For the model, Transformer Normalized Discounted Cumulative Gain (NDCG) result at top 10: 0.0849, Popular recommendation NDCG result at top 10: 0.0082. On the other hand, for rating model, we have precision @ 50: 0.8530, recall @ 50: 0.9402, RMSE: 1.0518. It should be noted that for Table 1, section of sequential nature, only one movie was used to predict the next few sequential values, while for Table 2, for instance, that value is changed to three.

<table>
<tr><td colspan="2">user_110007</td></tr>
<tr><th>Some of watched movies that the model used</th><th>Personalized Recommendation based on rating</th></tr>
<tr><td>['Dear Diary (Caro Diario) (1994) - Comedy|Drama', 'Ghost (1990) - Comedy|Drama|Fantasy|Romance|Thriller', 'Nelly & Monsieur Arnaud (1995) - Drama', ' Grass Harp, The (1995) - Comedy|Drama', 'Third Man, The (1949) - Film-Noir|Mystery|Thriller', 'Deer Hunter, The (1978) - Drama|War', 'Down by Law (1986) - Comedy|Drama|Film-Noir', 'Sprung (1997) - Comedy', 'Switchback (1997) - Crime|Mystery|Thriller']</td><td>-Star Wars: Episode VI - Return of the Jedi (1983)<br>-Raiders of the Lost Ark (Indiana Jones and the Raiders of the Lost Ark) (1981)<br>-Princess Bride, The (1987)<br>-E.T. the Extra-Terrestrial (1982)<br>-Seven (a.k.a. Se7en) (1995)<br>-Back to the Future Part II (1989)<br>-Godfather, The (1972)<br>-Mission: Impossible (1996)<br>-Twister (1996)<br>-Indiana Jones and the Last Crusade (1989)<br>-Babe (1995)</td></tr>
<tr><td colspan="2">Account for sequential nature of the watched movies</td></tr>
<tr><th>watched movies that the model used</th><th>Personalized Recommendation based on sequence of watched movies</th></tr>
<tr><td>-Back to the Future (1985)</td><td>-Lion King, The (1994)<br>-Star Wars: Episode VI - Return of the Jedi (1983)<br>-Raiders of the Lost Ark (Indiana Jones and the Raiders of the Lost Ark) (1981)<br>-Princess Bride, The (1987)<br>-E.T. the Extra-Terrestrial (1982)<br>-Seven (a.k.a. Se7en) (1995)<br>-Back to the Future Part II (1989)<br>-Godfather, The (1972)<br>-Mission: Impossible (1996)<br>-Twister (1996)<br>-Indiana Jones and the Last Crusade (1989)<br>-Babe (1995)<br>-Toy Story (1995)<br>-Terminator, The (1984)<br>-Groundhog Day (1993)<br>-Pretty Woman (1990)<br>-Gone with the Wind (1939)<br>-Speed (1994)<br>-Star Wars: Episode IV - A New Hope (1977)</td></tr>
</table>

Table 1: Table The first section based on rating dataset, related to user with ID of 110007

### 3.0.2 Discussion

Effective recommendation systems often leverage more than a single technique, e.g., ratings and sequences. So, to account for the sequential historical interaction of users with different movies, we incorporated sequential users' behaviors for watched movies, in addition to rating.

A hybrid approach can provide a comprehensive understanding of user preferences and generate more accurate recommendations. In this study, we used collaborative filtering, using two approaches of rating and sequential nature of the watched movies to spot a similar taste across similar users. The technique could be used to take advantage of two or more techniques, while minimizing the limitations of individual ones.

In the next paragraphs we will discuss some of the techniques that we could ultimately use to aggregate the results of the two techniques. Weighted averaging is a possibility where we could assign weights to the employed recommendation technique, based on their perceived importance or the expert opinion. Thresholding is another consideration, where we might set thresholds for each method and only consider movies that meet or exceed these thresholds. Afterward, we might combine the results by averaging. Then, after setting thresholds, the upcoming techniques could be based on weighted averaging again.

Another option is a hybrid method where a machine learning model could be used to learn how to best combine the results from both techniques. Input features could include the individual scores from each technique, and the model

<table>
<tr><td>user_65543<br>Rating (Already watched movies)</td><td>Personalized Recommendation with Transformer</td></tr>
<tr><td>['Dangerous Minds (1995) - Drama',<br>'Just Cause (1995) - Mystery|Thriller',<br>'Barb Wire (1996) - Action|Sci-Fi',<br>'American Dream (1990) - Documentary',<br>'Bob Roberts (1992) - Comedy',<br>'Armour of God (Long xiong hu di) (1987) - Action|Adventure|Comedy',<br>'Fire Within, The (Feu follet, Le) (1963) - Drama',<br>'Maniac (1980) - Horror',<br>'Little Giants (1994) - Children|Comedy',<br>'Comedy of Terrors, The (1964) - Comedy|Horror']</td><td>-Billy Elliot (2000)<br>-State and Main (2000)<br>-Beverly Hills Cop (1984)<br>-Scary Movie (2000)<br>-Bridget Jones's Diary (2001)<br>-Tootsie (1982)<br>-Wonder Boys (2000)<br>-Whole Nine Yards, The (2000)<br>-City Slickers (1991)<br>-Chocolat (2000)</td></tr>
<tr><td>watched movies that the model used</td><td>Personalized Recommendation based on sequence of watched movies</td></tr>
<tr><td>Input Sequence:<br>-Planes, Trains & Automobiles (1987)<br>-Commitments, The (1991)<br>-Fletch (1985)</td><td>Recomendations:<br>-Billy Elliot (2000)<br>-State and Main (2000)<br>-Beverly Hills Cop (1984)<br>-Scary Movie (2000)<br>-Bridget Jones's Diary (2001)<br>-Tootsie (1982)<br>-Wonder Boys (2000)<br>-Whole Nine Yards, The (2000)<br>-Dirty Rotten Scoundrels (1988)<br>-Mummy Returns, The (2001)<br>-Naked Gun 2 1/2: The Smell of Fear, The (1991)<br>-Dude, Where's My Car? (2000)<br>-Unbreakable (2000)<br>-Jay and Silent Bob Strike Back (2001)<br>-Seven Year Itch, The (1955)<br>-City Slickers (1991)<br>-Chocolat (2000)<br>-Pay It Forward (2000)<br>-Nurse Betty (2000)</td></tr>
</table>

Table 2: Table The second section based on rating dataset, related to user with ID of 65543

could be trained to output a final recommendation score. Those labeled observations work as our training data, helping the machine learning model learn which recommendations are best.

In summary, as we are interested in combining two recommendation systems, after setting c=2, the prediction score follows:

$$p_{u,i} = \sigma_1 . p_{u,i}^{(1)} + (1 - \sigma_1) . p_{u,i}^{(2)} \tag{8}$$

Where $\sigma_f$ denotes weights of algorithm, and $p_{u,i}$ highlights prediction score for user *u* for item *i*. As we will obtain both techniques separately, we could apply the above equations separately and based on the discretion of the experts.

In the current study, in the rating part of the recommendation system, we used explicit feedback left by the users. The system is designed in a way to predict and suggest items that are aligned with users' interests, where the systems is based on presenting items based on users' past behaviors and those items that the users might prefer in the future.

Challenges remain in the form of data-scarcity, cold-start, time consumption, scalability, and accuracy. To address those challenges various techniques could be employed, such as demographic based recommendation system (DBRS), where it provides recommendations to users based on their demographic characteristics, e.g., gender, age, locations, etc.

### 3.0.3 Study imitations

Although, in this study, recommending relevant movies was not directly based on favorite genre, actor or director, it is expected the methodological approach accounted for those confounding factors. In future study, both explicit and implicit data might be employed. That is by gaining insights based on history of the items watched by users' feedback domain knowledge and users' demographic characteristics. Those will help to better identify the items that best fit the users' needs and preferences. Also, in addition to hybrid technique, fuzzy expert systems could be employed using a set of predefined if-then rules created by domain experts. These rules are designed to handle imprecision and uncertainty by assigning degrees of truth to different conditions.

Knowledge-based recommender system (KBRS) is another way to enhance the system significantly after gaining more insights about the problem. KBRS works based on the domain knowledge about the items and users. The technique is based on reasoning on what items are related to the users' interests [13]. So, the technique might be suitable for cases when the items are not approached regularly by users. On the other hand, the main limitations of KBRS is the requirement of knowledge and engineering skills [7].

Lack of sufficient rating might impact providing trustworthy collaborative filtering, which might be aligned with cold start. To address that issue, collaborative filtering could be accompanied by content-based filtering. CF could perform especially well when there are so many interactions between users and the items, otherwise we will have a very sparse matrix. One solution for those scenarios is to use implicit review or use of content-based filtering.

Application of sequential data, in addition to rating, is especially beneficent as rating technique alone might not provide recommendations on items where there is no rating. To address that drawback, collaborative filtering and in-content filtering could be combined, being another type of hybrid technique. Weighted technique could use hybrid technique where the prediction score is the combination linear of score of the combined technique.

We discussed in the method section that various weights based on the expert opinion would be assigned to the identified movies based on the two techniques and the final movies would be proposed based on movies with highest weights. However, we left those weights unknown as those would be made known based on expert opinion. For instance, as sequential nature of the datasets considers both like and disliked movies, a higher weight might be assigned to the rating types of recommendations techniques as only movies with a better rating were extracted but the opinion of the watchers regarding the sequential movies are not necessarily favorable. It should be noted that in real-life problem explicit rating might not be available for all users and separate rule-based algorithms might be designed for those users leaving the rating by assigning a higher weight to those rating and lower weighting to the sequential nature of the dataset and for those users who did not leave any rating, the sequential nature of the dataset might be used solely.

another limitation of our approach is that we opted not to employ a Two-Tower architecture, which could have provided a more explicit representation of user-item interactions by maintaining distinct embedding spaces for users and items. A Two-Tower model might have enhanced feature extraction for personalization, particularly in capturing nuanced relationships between users and items. However, our choice to implement a weighted fusion method instead allowed for greater flexibility in balancing rating-based and sequential collaborative filtering approaches.

While a Two-Tower approach typically requires well-structured embeddings for both users and items, our potential technique might dynamically adjusts its reliance on ratings or sequential viewing history. This adaptability makes it more suitable for cases where one type of data is sparse or inconsistent. By optimizing weighting schemes based on data availability, our method remains robust in scenarios with uneven distributions of rating and sequential data, ensuring effective personalization without requiring extensive feature engineering.

#### 3.0.4 Conclusion

In conclusion, this study underscores the pivotal role of hybrid recommendation systems in addressing the challenges of information overload by leveraging both collaborative filtering approaches—sequential behavior analysis and movie ratings. By introducing a weighted post-fusion methodology, we demonstrated how the dynamic adjustment of weights based on specific use cases can enhance the accuracy and relevance of recommendations. Our findings emphasize the importance of integrating sequential data and rating information to capture both short-term and long-term user preferences, particularly in scenarios where standalone models fall short. This approach not only provides a more nuanced understanding of user preferences but also lays the groundwork for future research in personalized recommendation systems.

#### 3.0.5 Appendix

#### 3.0.6 Sequential analysis, process summary: method

Here we outline the process/steps of converting the watched movies in a format of sequential analysis.

- To prevent the IDs from being written as integer or float data type, we attached user_ to the users IDs and attached movie_ to the movies.
- Assign numerical indices into unique ID to various users and movies.
- Sort the data based on the timestamp and grouped them based on user_ids before creating sequences of the watched movies.
- After sorting the movies and grouping them based on user_id, organize them based on step-size of 2 and sequence/window length of 4, so the list is created by using start_index + window size which is the length of each sequence. The start_index is then incremented by step_size to slide the window forward.
- Test and train were split by a ratio of 80 and 20 percent, while the shuffling of the data was controlled by random_state, then the data was filtered to include only user_id and sequence of movies ID for columns of the data.
- Then, we collate and load the data by dataloader based on assigned batch size of 256.
- The transformer that we used for analyzing the sequence of movie is naïve, which processes all input tokens in parallel, and would not account for the order of the sequence. So, to account for the movie sequence and its order, we used positional encoder, which is based on cosine and sin, which then would be added to the original values.
- The model, which we here call transformerneural model, is a core component of the model architecture being used for recommendation system. It initialized positional encoder, a multihead attention-based transformer encoder, and embeddings of movies and users. During the forward pass, we embed the movie and user IDs, and applied positional encoding to movie embeddings and process them through the transformer encoder and concatenate the user embedding with the transformer layer. Finally we used a linear layer for mapping the combined embeddings to the movie vocabulary for generating the final output.
- Next, training would be done by reshaping the data and splitting movies sequences into inputs and targets, where the input are the sequence without the last element and the target are the sequence without the first element, and then predicting the movies sequences, calculating the loss and updating the model parameters through backpropagation.
- Cross-entropy loss was used as the criterion for calculating loss and performing backpropagation. Stochastic Gradient Descent (SGD) was employed as the optimization function.
- top_10_movies were used as the baseline recommendation system to be compared with our designed technique. The technique is based on movies with the highest number of watched. Where we used those movies with the highest 95% quantile for the count of the watched movies, where we obtained the means of rating across all movies and finally used a weighted rating that takes into account both the average rating and the number of ratings that helps us to balance the popularity and the quality of the movies. The formula for the weighted considers both average rating for the movies, number of votes and minimum required number for a movie to be included based on $95^{th}$ percentile or minimum of required votes for a movie to be listed.
- Normalized discounted cumulative gain (NDCG), as a ranking quality metric, was used in the system to evaluate the quality of a ranked list of items and compare it with ranking produced by the model. To do so, we iterate over validation/test dataset, reshape the output and obtained the predicted/recommended 10 top movies.
- For each batch of movie and user data, the model predicts the next *k* movies in the sequence. The top predictions are filtered to exclude already-watched movies, and the results are stored in transformer_reco_results. Additionally, the last watched movie is decoded and checked against a list of popular movies, with the results stored in popular_reco_results. The process aims to ensure that recommended movies are new to the user and checks their popularity. The final output is a list of transformer-based recommendations and popular movie recommendations. We check if the last watched movie is among the top recommendations or popular movies for coming up with popular_reco_result. Finally, NDCG was used to come up with the results.

- Finally, we generate recommendations based on the trained model by feeding movie and user id to the model. The function takes a user ID and a sequence of watched movies, tokenizes and encodes them, and then passes these tensors through a pre-trained model to predict the next movies that users likely to watch. It uses torch.no_grad() to perform inference without tracking gradients, retrieves the top *k* probable movies, and filters out already watched ones to return a list of recommended movie titles.

The next few paragraphs highlight the processes for rating analysis.

### 3.0.7 Rating analysis: method

For this type of analysis, instead of using sequential nature of movies, we solely focus on the rating associated with each movie. Like the previous model architect, this model also uses embedding layers for users and movies as embeddings. As we discussed they are a powerful tool in recommendation systems in providing a way to transform categorical data (like user IDs and movie IDs) into a continuous space. This continuous representation allows the model to capture and learn the patterns in the data.

Similar processes as before were implemented where the few paragraphs will discuss the details.

- The MovieLensDataset class, as a custom PyTorch Dataset, was used to handle user, movie, and rating data, being initialized with three lists: users, movies, and ratings.

- We then perform the following steps by first generating embeddings, including embeddings for users and movies. We then concatenate the embeddings. This combination is a crucial step as it merges user and movie information into a unified representation. We then pass that through Hidden Layers and the combined data then passes through fully connected layers with ReLU activation and dropout. These layers are where the model learns to predict ratings based on the combined user-movie representation.
- To set up the recommendation system model in PyTorch, we first define the loss function using nn.MSELoss(). Next, we initialize the optimizer with torch.optim.Adam, and setting the learning rate to 1e-3 and passing the model parameters. In the recommendation model itself we specify the number of users and movies based on the length of the label encoder classes, an embedding size of 64, a hidden dimension of 128, and a dropout rate of 0.1.
- To evaluate the recommendation model, we first set it to evaluation mode, ensuring no changes in weights are applied. Then, we iterate through the validation data loader. For each batch, we predict the ratings by passing user and movie data to the model and compare these predictions to the ground truth ratings. The predicted and actual ratings are stored in y_pred and y_true lists, respectively. After collecting all predictions, we calculate the Root Mean Squared Error (RMSE).
- A function computes precision and recall for a list of user ratings. That is done by first sorting the ratings in descending order based on the estimated rating to make sure the top-K items are the ones with the highest estimated ratings. It then calculates the number of relevant items and the number of recommended items in the top-K list. The function also determines the number of items that are both relevant and recommended in the top-K list. Precision is calculated as the proportion of relevant items among the recommended items, the relevant was considered for items with at least rating of 3, while recall is the proportion of relevant items among all relevant items.
- The calculate_ndcg function computes the NDCG for a list of user ratings. It first sorts the ratings in descending order based on the estimated ratings. Then, it extracts the true ratings and predicted scores, reshaping them to fit the expected input format for the ndcg_score function. Finally, it calculates and returns the NDCG value for the top-K items, which measures the ranking quality of the recommendations.
- A function was built to generate personalized movie recommendations for a user by leveraging a trained model. It first filters out movies the user has already seen, then predicts ratings for the remaining unseen movies in batches. Using the model, it computes predicted ratings for each batch and sorts them in descending order. At the end, it returns the top-K movies with the highest predicted ratings, which will provide the user with a list of recommended movies.